\documentclass[12pt]{article}
\usepackage[utf8]{inputenc}
\usepackage[dvips, margin=1in]{geometry}
\usepackage{bm}
\usepackage{amsmath, amssymb, amsthm}
\usepackage{enumitem}
\usepackage{color}
\usepackage{float}
\usepackage{graphicx}
\usepackage{caption}
\usepackage{subcaption}
\usepackage{multicol}
\RequirePackage[authoryear]{natbib}
\usepackage{authblk}
\usepackage{lineno}
\usepackage{setspace}
\usepackage{hyperref}
\hypersetup{
    colorlinks=true,
    citecolor=blue,
    linkcolor=blue,
    urlcolor=blue
}
\begin{document}

\title{\bf Multi-scale dynamic latent space models improve inference for a social predator}
\author[1]{Xinyi Lu\thanks{Corresponding author. Email: lucy.lu@usu.edu.}}
\author[2]{T.J. Clark-Wolf}
\author[3]{Matthew C. Metz}
\affil[1]{\small\textit{Mathematics and Statistics Department, Utah State University, Logan, UT, USA}}
\affil[2]{\small\textit{Department of Wildland Resources and Ecology Center, Utah State University, Logan, UT, USA}}
\affil[3]{\small\textit{Yellowstone Center for Resources, Yellowstone National Park, WY, USA}}
\date{}
\maketitle

\section*{Abstract}
Social networks are dynamic in nature. Examining changes in dynamic networks provides insights into the drivers and consequences of social interactions. Dynamic latent space models (DLSMs) quantify social tie strength through a distance measure between latent positions in an unobservable, low-dimensional space. Conditional on the latent positions, ties are assumed independent, and temporal dynamics are introduced through autocorrelated trajectories. These trajectories are typically modeled as random walks; however, random walks lack the mechanism to characterize network clustering, where individuals from the same group tend to form stronger ties than those across groups. Therefore, we extend the framework of DLSM to distinguish within-group dynamics from between-group dynamics. Our approach employs coupled movement models operating at two scales in the latent space to encode community structure using autoregressive processes with attraction. Through simulation studies, we show that when cluster separation is high, our framework improves coverage of linear predictors and enhances overall model fit relative to a naive DLSM. When cluster separation is moderate, performance remains comparable to that of the naive approach. In addition, our method demonstrates empirical computational advantages over the naive DLSM, with potential for further efficiency gains. We apply our framework to a case study of gray wolf (\textit{Canis lupus}) interactions over three decades in Yellowstone National Park, where it provides improved model performance and novel insights into wolf social dynamics.
\bigbreak
Key words: gray wolf, multi-scale decomposition, state-space model, statistical network analysis, Yellowstone National Park

\section{Introduction}\label{intro}
Social systems exhibit complex patterns of interactions \citep{sumpter2010collective}, and network-based representations provide a principled framework for quantifying the development of social interactions \citep{krause2002living}. Social networks are inherently dynamic, evolving through changes in relationships among individuals and by individuals joining or leaving the network \citep{pinter2014dynamics}. Examining these changes offers an understanding of the drivers and consequences of social interactions \citep{whitehead2008analyzing}. However, modeling dynamic network data remains challenging due to methodological and computational difficulties. \citep{fisher2017analysing, finn2019use}.
\bigbreak
Conventional statistical tools for dynamic network analysis are represented by the Stochastic Actor Oriented Model (SAOM) \citep{frank1986markov, snijders2001statistical} and the Temporal Exponential Random Graph Model (TERGM) \citep{hanneke2006discrete, leifeld2018temporal}. Both methods are rooted in the (static) Exponential Random Graph Model (ERGM) \citep{wasserman1996logit}, where the observation unit is an individual and intrinsic dependence is quantified by a counting measure defined on network permutations. These methods are useful for identifying network substructures such as reciprocity, transitivity, and homophily. The accessibility of SAOM through open-sourced softwares (e.g., \texttt{RSiena} \citep{ripley2025manual}) and its ability to model co-evolution between network and individual behaviors \citep{snijders2018beyond} have made it popular across a wide range of applications \citep{borgeaud2016intergroup, fisher2017analysing}. However, SAOM can be computationally prohibitive for high-dimensional networks \citep{shappell2025accounting} and is fundamentally limited to modeling discrete network data. Most of its applications use binary data, including cases where continuous data were dichotomized to accommodate the framework \citep{steglich2006applying, ilany2013structural}. Simplifying rich network data for model fitting can limit the insights gained relative to the effort and resources invested, and ignoring interaction weights can inflate uncertainty in parameter estimation, consequently increasing type-II error rate in hypothesis testing \citep{franks2010sampling, farine2014measuring}. 
\bigbreak
An alternative framework for dynamic network analysis is the Dynamic Latent Space Model (DLSM) \citep{sewell2015latent, sewell2016latent}. DLSM is rooted in the Latent Space Model (LSM) \citep{hoff2002latent}, where the observation unit is a tie (between a pair of individuals) and intrinsic dependence is quantified by a distance measure defined on a latent ``social" space. The strength of a tie is reflected by the inferred proximity of the pair in the latent space: Closer positions indicate stronger ties. Because the distance measure is a continuous measure, it is applicable to both discrete and continuous network data. DLSM specifies a sequence of autocorrelated positions in the latent space, culminating in a latent trajectory for each individual. Mechanistic latent trajectory models (i.e., movement models in latent space) are capable of characterizing complex spatio-temporal dynamics \citep{lu2023latent, lu2025regularized}, yet there are few applications of advanced movement models beyond a random walk in a DLSM. \cite{sewell2015latent} proposed a framework to detect edge attraction similar to collective movement models that characterize intra-group interactions \citep{brillinger1998elephant, scharf2016dynamic}. However, the analysis was conducted post-hoc using the outputs from a random walk DLSM.
\bigbreak
Our approach extends the DLSM by explicitly distinguishing intra- and inter-group dynamics under a unified framework informed by network clustering. Clustering is typical of social networks where stronger interactions occur between individuals from the same group than those across different groups. The groups may be observed as an explanatory variable or inferred from the data. \cite{holland1983stochastic} and \cite{wang1987stochastic} represent the first works to explore clustered networks using the Stochastic Block Model (SBM). Since then, various extensions have been developed for weighted networks \citep{nowicki2001estimation, karrer2011stochastic} and dynamic networks \citep{fu2009dynamic, xing2010a}). A fundamental discrepancy between an SBM and a DLSM lies in intrinsic dependence: In an SBM, the ties are assumed independent conditioning on group membership (i.e., stochastic equivalence \citep{nowicki2001estimation}). We relax this assumption through coupled movement models defined on two distinct scales in the latent space. We model inter-group dynamics using random walks at a coarse latent spatial scale to capture highly variable interactions, and intra-group dynamics using autoregressive processes with attraction at a finer scale to represent group cohesion. This decomposition not only facilitates scale-specific inference and cross-scale comparisons \citep{lu2024using} but has also been shown to enhance computational efficiency in various contexts. \citep{dogra2017multi, cao2022scalable}. Variations of DLSM have been applied to network clustering \citep{sewell2017latent, wang2025dynamic}; however, unlike our proposed framework, these approaches operate on the assumptions of a fixed number of groups and group-specific parameters that are time-homogeneous. In contrast, our framework accommodates emerging or disappearing groups as well as time-varying inter-group relationships. This flexibility allows our framework to realistically represent highly transient social, ecological, or organizational systems.
\bigbreak
We present our Multi-Scale Dynamic Latent Space Model (MS-DLSM) in Section \ref{method}. In Section \ref{sim}, we demonstrate the framework using simulated examples. In Section \ref{case}, we demonstrate the framework using network data on gray wolves (\textit{Canis lupus}) in Yellowstone National Park. Finally, in Section \ref{diss}, we discuss practical recommendations and possible extensions of our framework.

\section{Method}\label{method}
\subsection{Data Model}\label{data}
Using terminology from statistical network analysis, we refer to an individual in a social network as a ``node" and the relationship between two distinct nodes an ``edge". A network is uniquely defined by its composing set of nodes and edges. In a dynamic network, the nodes and edges change over time, and we denote $\mathcal{S}_t \equiv\{\mathcal{N}_t,\ \mathcal{Y}_t\}$ the observed network at time $t$, $t = 1, \dots, T$, where $\mathcal{N}_t$ is the composing set of nodes and $\mathcal{Y}_t$ the set of edges. We denote $y_{ijt}$, an element of $\mathcal{Y}_t$, the observed edge between nodes $i$ and $j$ at time $t$, where $i, j\in \mathcal{N}_t$ and $i \neq j$. For continuously weighted edges between 0 and 1, we define a logit-linear mixed model for the expected edge weights,
\begin{align}
    \text{logit}\left(y_{ijt}\right) &\equiv \log\left(\frac{y_{ijt}}{1 - y_{ijt}}\right) =  \bm{x}_{ijt}'\bm{\beta} - \log(d_{ijt}) + \epsilon_{ijt},\label{y}\\
    \epsilon_{ijt}\left|\sigma^2\right. &\sim \text{N}\left(0,\ \sigma^2\right),\label{e}
\end{align}
where $\bm{x}_{ijt}$ is a vector of explanatory variables and $\bm{\beta}$ their associated effects. Variable $d_{ijt}$ is the latent distance between $i$ and $j$ which we elaborate in Section \ref{process}. The effect of log distance is fixed at -1 for model identifiability. The strength of an edge between two nodes is therefore inversely related to their distance in the latent space. Lastly, $\epsilon_{ijt}$ is Gaussian random noise that accounts for measurement error.
\bigbreak
The above model is readily adaptable to various data types including binary and count edges \citep{sewell2015latent, sewell2016latent}. While it is specified for undirected networks in accordance to our case study, we discuss the extension of our method to directed networks in Section \ref{diss}.  

\subsection{Dynamic Latent Space Model}\label{process}
The latent distance $d_{ijt}$ functions as a random effect in a generalized linear mixed model that accounts for variation in $y_{ijt}$ unexplained by $\bm{x}_{ijt}$. We emphasize that $d_{ijt}$ is fundamentally different from the derived distance in state-space movement models \citep{patterson2008state, auger2021guide}. These models from the animal movement literature typically operate in a geographic space, where true animal locations are considered latent due to observation uncertainty (e.g., precision of GPS data). There are studies that relate the latent, albeit geographic distance between animals to their social relationships \citep{jacoby2016inferring, scharf2016dynamic}. However, our model does not use any location data, and our distance measure is defined in a social space rather than a geographic space.
\bigbreak
To account for the contrast in edge propensity in a clustering network, we formulate two distinct movement processes that contribute to $d_{ijt}$: One on the node scale (i.e., intra-group) and one on the cluster scale (i.e., inter-group). We denote $\bm{z}(i, t)$ the position of node $i$ at time $t$ in the latent space. For a two-dimensional space, $\bm{z}(i,t) \equiv (z_1(i, t),\ z_2(i, t))'$. We also denote $c(i, t)$ the observed group membership of node $i$ at time $t$, where $c(i, t)\in\mathcal{C}_t$, the set of observed groups at time $t$. Lastly, we denote $\bm{\mu}(c(i, t))$ the movement center associated with $c(i, t)$.
\bigbreak
At the node scale, we specify an autoregressive process model with attraction toward the group-affiliated movement center. The attraction mechanism regularizes individual latent trajectories toward the center, thereby capturing within-group cohesion. The position of node $i$ at $t_0(i)$, its time of first observation, is Gaussian distributed with isotropic variance around its concurrent center,
\begin{equation}\label{i1}
    \bm{z}(i, t_0(i))\left|\bm{\mu}(c(i, t_0(i))), \sigma^2_{z_0}\right. \sim \text{N}\left(\bm{\mu}(c(i, t_0(i))),\ \sigma^2_{z_0}\textbf{I}\right).
\end{equation}
For every time after $t_0(i)$ until its last observation at $T(i)$, the trajectory of node $i$ follows a first-order autoregression with correlation $\rho\in(0, 1)$,
\begin{equation}\label{i2}
    \bm{z}(i, t)\left|\bm{z}(i, t-1), \bm{\mu}(c(i, t)), \bm{\mu}(c(i, t-1)), \rho, \sigma^2_z\right. \sim \text{N}\left(\rho\bm{z}(i, t-1) + \bm{\mu}(c(i, t)) - \rho\bm{\mu}(c(i, t-1)),\ \sigma^2_z\textbf{I}\right).
\end{equation}
In the case of a constant movement center and no group change (i.e., $\bm{\mu}(c(i, t)) = \bm{\mu}^*$ for $t = t_0(i), \dots, T(i)$), the resulting process reduces to 
$$
\bm{z}(i, t)\left|\bm{z}(i, t-1), \bm{\mu}^*, \rho, \sigma^2_z\right.\sim \text{N}\left(\rho\bm{z}(i, t-1) + (1 - \rho)\bm{\mu}^*, \sigma^2_z\textbf{I}\right)
$$
and becomes second-order stationary. Nonetheless, stationarity is not generally guaranteed nor necessary in our framework to accommodate group change and inter-group dynamics through the cluster movement process that we describe below. 

\bigbreak
At the cluster scale, we simplify the notation $\bm{\mu}(c(i, t)) \equiv \bm{\mu}(c, t)$ for all nodes whose group membership is $c$ at time $t$. We specify the group-affiliated center at $t_0(c)$, its time of first observation, to be Gaussian distributed with isotropic variance around the origin,
\begin{equation}\label{c1}
    \bm{\mu}(c, t_0(c))\left|\sigma^2_{\mu_0}\right. \sim \text{N}\left(\bm{0},\ \sigma^2_{\mu_0}\textbf{I}\right).
\end{equation}
For every time after $t_0(c)$ until its last observation at $T(c)$, the center affiliated with group $c$ follows a random walk,
\begin{equation}\label{c2}
    \bm{\mu}(c, t)\left|\bm{\mu}(c, t-1), \sigma^2_\mu\right. \sim \text{N}\left(\bm{\mu}(c, t-1),\ \sigma^2_\mu\textbf{I}\right). 
\end{equation}
A realization from this process is a mean-zero trajectory with variance that increases over time. This process is well suited for social systems with weak between-group interactions. Groups are assumed to originate from a common region of the latent space, implying that newly formed groups initially occupy nearby social positions. As additional groups emerge and the number of groups grows, the increasing variance of the process promotes greater separation among existing groups. This mechanism captures the idea that expanding group structure requires increasing social spacing, leading to progressive dispersion of latent group positions over time.
\bigbreak
The joint posterior distribution of the above hierarchical model is
\begin{align*}
&\left[\bm{\beta}, \sigma^2, \{\bm{z}(i, t)\}, \sigma^2_z, \sigma^2_{z_0}, \rho, \{\bm{\mu}(c, t)\}, \sigma^2_\mu, \sigma^2_{\mu_0}\middle|\bm{Y}\}\right] \propto \prod_{t = 1}^T\prod_{(i, j)\in\mathcal{Y}_t}\left[y_{ijt}\middle|\bm{\beta}, \bm{z}(i, t), \bm{z}(j, t), \sigma^2\right] \times \\
& \prod_{i = 1}^n \left\{\prod_{t = t_0(i)+1}^{T(i)}\left[\bm{z}(i, t)\middle|\bm{z}(i, t-1), \bm{\mu}(c(i, t)), \bm{\mu}(c(i, t-1)), \rho, \sigma^2_z\right] \times \left[\bm{z}(i, t_0(i))\middle|\bm{\mu}(c(i, t_0(i))), \sigma^2_{z_0}\right]\right\} \times\\
& \prod_{c = 1}^C\left\{\prod_{t = t_0(c)+1}^{T(c)}\left[\bm{\mu}(c, t)\middle|\bm{\mu}(c, t-1), \sigma^2_\mu\right] \times \left[\bm{\mu}(c, t_0(c))\middle|\sigma^2_{\mu_0}\right]\right\} \times \\
&[\bm{\beta}] \times \left[\sigma^2\right] \times [\rho] \times \left[\sigma^2_z\right] \times \left[\sigma^2_{z_0}\right] \times \left[\sigma^2_\mu\right] \times \left[\sigma^2_{\mu_0}\right],
\end{align*}
where $\bm{Y}$ is the array of all observed edges, $\{\bm{z}(i, t)\}$ is the set of all latent nodal positions, $\{\bm{\mu}(c, t)\}$ is the set of all latent movement centers, $n$ is the total number of unique nodes, and $C$ is the total number of unique groups.

\subsection{Multi-scale Decomposition of Latent Distance}\label{ms}
DLSMs typically quantify $d_{ijt}$ using the $L_p$-norm of displacement between nodes $i$ and $j$ at time $t$, where $p$ is the latent space dimension \citep{hoff2002latent, sewell2015latent}. Following Section \ref{process}, this can be calculated using $\bm{z}_{i, t}$ and $\bm{z}_{j, t}$ from the nodal movement model (Equations \ref{i1} and \ref{i2}). Nonetheless, in highly clustered networks where inter-group associations are minimal, decomposing the distance matrix at time $t$ by groups and approximating inter-group displacements using movement centers (Equations \ref{c1} and \ref{c2}) can enhance computational efficiency. Consider a network at time $t$ with $n$ nodes and $C$ groups; assuming the groups are equally sized yields $n/C$ nodes per group. Calculating all pairwise distances costs $\mathcal{O}\left(\frac{n(n-1)}{2}\right)$. Replacing distances of inter-group pairs using the distance of their movement centers reduces the cost to $\mathcal{O}\left(\frac{n(n/C-1)}{2} + \frac{C(C-1)}{2}\right)$, whose magnitude depends on $C$. Under the assumption of conditional independence, we can compute intra-group distances in parallel across all groups, which further reduces the cost to $\mathcal{O}\left(\frac{n(n/C-1)}{2C} + \frac{C(C-1)}{2}\right)$. 
Therefore, we propose a decomposition of the latent distance,
\begin{equation}\label{d}
    d_{ijt} = \begin{cases}
        ||\tilde{\bm{z}}(i, t) - \tilde{\bm{z}}(j, t)||_p,\ c(i, t) = c(j, t)\\
        ||\bm{\mu}(c(i, t)) - \bm{\mu}(c(j, t))||_p,\ c(i, t) \neq c(j, t)
    \end{cases},    
\end{equation} 
\bigbreak
with the following marginalized nodal movement model
\begin{align}
    \tilde{\bm{z}}(i, t_0(i))\left|\sigma^2_{z_0}\right. &\sim \text{N}\left(\bm{0},\ \sigma^2_{z_0}\textbf{I}\right),\\\label{i1t}
    \tilde{\bm{z}}(i, t)\left|\tilde{\bm{z}}(i, t-1), \rho, \sigma^2_z\right. &\sim \text{N}(\rho\tilde{\bm{z}}(i, t-1),\ \sigma^2_z\textbf{I}),\ t = t_0(i) + 1, \dots, T(i).
\end{align}
In addition to reducing computational burden, marginalized likelihood has also been shown to improve estimation performance in high-dimensional parameter settings \citep{ver2024marginal}. Therefore, the joint posterior distribution of our proposed MS-DLSM becomes
\begin{align*}
&\left[\bm{\beta}, \sigma^2, \{\tilde{\bm{z}}(i, t)\}, \sigma^2_z, \sigma^2_{z_0}, \rho, \{\bm{\mu}(c, t)\}, \sigma^2_\mu, \sigma^2_{\mu_0}\middle|\bm{Y}\}\right] \propto \\
&\prod_{t = 1}^T\left\{\prod_{\substack{(i, j)\in\mathcal{Y}_t\\ c(i, t)=c(j, t)}}\left[y_{ijt}\middle|\bm{\beta}, \tilde{\bm{z}}(i, t), \tilde{\bm{z}}(j, t), \sigma^2\right] \times \prod_{\substack{(i, j)\in\mathcal{Y}_t\\ c(i, t)\neq c(j, t)}}\left[y_{ijt}\middle|\bm{\beta}, \bm{\mu}(c(i, t)), \bm{\mu}(c(j, t)), \sigma^2\right]\right\} \times \\
& \prod_{i = 1}^n \left\{\prod_{t = t_0(i)+1}^{T(i)}\left[\tilde{\bm{z}}(i, t)\middle|\tilde{\bm{z}}(i, t-1), \rho, \sigma^2_z\right] \times \left[\tilde{\bm{z}}(i, t_0(i))\middle|\sigma^2_{z_0}\right]\right\} \times\\
& \prod_{c = 1}^C\left\{\prod_{t = t_0(c)+1}^{T(c)}\left[\bm{\mu}(c, t)\middle|\bm{\mu}(c, t-1), \sigma^2_\mu\right] \times \left[\bm{\mu}(c, t_0(c))\middle|\sigma^2_{\mu_0}\right]\right\} \times \\
&[\bm{\beta}] \times \left[\sigma^2\right] \times [\rho] \times \left[\sigma^2_z\right] \times \left[\sigma^2_{z_0}\right] \times \left[\sigma^2_\mu\right] \times \left[\sigma^2_{\mu_0}\right].
\end{align*}
\bigbreak
We provide the full conditional distributions of our model and the prior distributions for simulation and case study in Supplementary Materials. Non-informative priors are chosen for most parameters and we justify specific choices of informative priors in Section \ref{case}.

\section{Simulation}\label{sim}
We conducted simulation studies to compare our proposed MS-DLSM with a naive DLSM in terms of estimation properties, model fitness, and computation. The naive DLSM serves as a benchmark with the correctly specified distance model for the simulated data, whereas the MS-DLSM is misspecified due to the distance decomposition (Equation \ref{d}). We do not consider other clustering DLSMs because they lack the mechanism to represent inter-group temporal dynamics in the latent space. The naive DLSM was used to generate data following Equations \ref{y}, \ref{e}, \ref{i1}, and \ref{i2}; however, instead of a distance decomposition, the latent distance was defined directly as $d_{ijt} = ||\bm{z}_{i, t} - \bm{z}_{j, t}||_p$. We randomly assigned $n = 100$ nodes to $C = 5$ equally sized groups observed over $T = 5$ time points, assuming fixed group memberships, for a clear visual representation (Figure \ref{sim_ts}). The initial movement centers in a two-dimensional latent space were generated using Equation \ref{c1} and their subsequent trajectories were generated using Equation \ref{c2}. Conditioned on the movement centers, the initial nodal positions were generated using Equation \ref{i1} and their subsequent trajectories were generated using Equation \ref{i2}. Finally, edge outcomes $y_{ijt}$ were generated using the exact $L_2$ distance between nodes.
\bigbreak
When network clustering is strong, the MS-DLSM mechanism aligns closely with the underlying network dynamics. Even though the model is misspecified, the resulting information loss is minimal. As clustering weakens, however, the discrepancy between the model mechanism and the true network dynamics increases, resulting in a greater loss of information, especially regarding inter-group interactions. In our simulation design, the strength of network clustering is governed by the relative magnitude of nodal movement compared to group movement. To assess performance across different clustering regimes, we fixed $\sigma^2_\mu = \sigma^2_{\mu_0} = 1$ and varied $\sigma^2_z$ and $\sigma^2_{z_0}$ to create three cases with progressively weaker clustering,  
\begin{enumerate}
    \item[] Case 1: $\sigma^2_{z_0} = \sigma^2_z = 0.01$;
    \item[] Case 2: $\sigma^2_{z_0} = \sigma^2_z = 0.1$;
    \item[] Case 3: $\sigma^2_{z_0} = \sigma^2_z = 1$.
\end{enumerate}
In addition, we designed three covariate arrays, $\bm{X}_1,\ \bm{X}_2$, and $\bm{X}_3$, to represent common types of explanatory variables in network analysis. The elements of $\bm{X}_1$ were continuous and edge-homogeneous, representing time-varying conditions that influence overall network density. The elements of $\bm{X}_2$ were binary and time-homogeneous, representing homophily effects associated with fixed biological traits. The elements of $\bm{X}_3$ were continuous and heterogeneous across both edge and time, representing homophily driven by individual characteristics or social status that evolve over time. We set $\bm{x}_{ijt} = (x_{1t},\ x_{2ij},\ x_{3ijt})'$, $\bm{\beta} = (0.5, 1, -2)'$, and $\sigma^2 = 1$. We also set $\rho=0.7$ to represent moderate temporal dependence in individual movement. 
\bigbreak
We independently simulated twenty data sets for each of the three cases. Figure \ref{sim_ts} displays one representative realization from each case. For every simulated data set, we obtained posterior inference under both the MS-DLSM and the naive DLSM using Markov chain Monte Carlo (MCMC) algorithms implemented in R \citep{R}. Each chain was run for 100,000 iterations, with the first 40,000 iterations discarded as burn-in. The remaining samples were thinned at a rate of 1/10. 
The latent positions in DLSMs are not uniquely identifiable due to rotational and translational invariance; nonetheless, the corresponding pairwise distances are identifiable \citep{hoff2002latent, sewell2015latent}. Therefore, posterior samples of latent configurations were aligned via Procrustes transformation to a reference configuration for visual identifiability \citep{goodall1991procrustes}, and latent space estimation was assessed via derived pairwise distances rather than on the latent coordinates themselves.
\bigbreak
We first considered the inference quality of the fixed effects, $\bm{\beta}$ (Figure \ref{sim_infer}). Bias was calculated as the difference between the Bayes estimator under squared error loss (i.e., posterior mean) and the true parameter. Coverage was calculated as the proportion of simulations in which the 95\% credible interval contained the true parameter. When cluster separation was high (Case 1), neither model exhibited systematic bias; however, the MS-DLSM achieved the nominal coverage, whereas the naive DLSM produced higher-than-nominal coverage, indicating overly wide credible intervals. Under moderate cluster separation (Case 2), neither model showed systematic bias. The naive DLSM again produced higher-than-nominal coverage, while the MS-DLSM produced lower-than-nominal coverage for $\beta_1$ and $\beta_2$. This pattern was expected, as the MS-DLSM may underestimate variability in $\bm{\beta}$ by imposing conditional independence among inter-group interactions. When cluster separation was low (Case 3), the naive DLSM exhibited lower bias and higher coverage across all $\beta$ coefficients. 
\bigbreak
Next, we evaluated overall model performance (Figure \ref{sim_fit}). The continuous ranked probability score (CRPS) was calculated as the integrated squared difference between the cumulative distribution function (CDF) (Equations \ref{y} and \ref{e}) and the empirical CDF of $y_{ijt}$, averaged across observations within each simulated data set. The deviance information criterion (DIC) accounted for model complexity by penalizing the additional parameters in the MS-DLSM. Distance correlation (DCor) was calculated as Pearson’s correlation between the true pairwise distances and those derived from the inferred latent space. For the MS-DLSM, $d_{ijt}$ was obtained using Equation \ref{d}; for the naive DLSM, $d_{ijt}$ was obtained using the $L_2$ distance between inferred nodal positions. Lower values indicate better performance for CRPS (with zero as optimal) and DIC, while higher values indicate better performance for DCor (with one as optimal). When cluster separation was high (Case 1), the MS-DLSM outperformed the naive DLSM across all evaluation criteria. Under moderate cluster separation (Case 2), the two models exhibited similar performance. When cluster separation was low (Case 3), the naive DLSM outperformed the MS-DLSM. Notably, the performance variability in MS-DLSM decreased as its performance improved (from Case 3 to Case 1), indicating increasingly consistent advantage as network clustering strengthened; on the other hand, the naive DLSM showed little change in performance variability across cases. While clustering strength in the simulation study was controlled by definition of $\sigma^2_z$ and $\sigma^2_\mu$, in practice, we propose a guideline for evaluating clustering strength by comparing observational data with simulated data, as described in Section \ref{diss}.
\bigbreak
Lastly, we examined computational cost (Figure \ref{time}). We varied the number of groups, $C$, while holding the total number of nodes fixed at $n = 100$ and maintaining equal group sizes. The MS-DLSM required less time to complete 1,000 MCMC iterations than the naive DLSM when the number of groups was fewer than fifty. However, as the number of groups increased and the number of nodes per group decreased, the computational cost of sampling from the group movement model outweighed the efficiency gained from distance decomposition. The computational advantage of the MS-DLSM was most pronounced at intermediate group sizes: With ten groups, the MS-DLSM (13s) was nearly twice as fast as the naive DLSM (22s).




\section{Case Study}\label{case}
\subsection{Data}
We applied our MS-DLSM to investigate interaction patterns among gray wolves (\textit{Canis lupus}) in Yellowstone National Park. 
Following their extirpation in the early 20th century, wolves were reintroduced to Yellowstone National Park in 1995 \citep{fritts1997planning}. Since reintroduction, the population has been intensively monitored through aerial tracking of radio-collared wolves during winter. During each flight, groups of wolves were recorded as those in close spatial proximity to one another. All pairs of radio-collared wolves that were within a group were assigned as having an interaction \citep{stahler2020ecology}. These interactions were aggregated over a winter-year (October--March) and standardized using the twice-weight index (TWI) to account for variation in detection frequency and observation effort \citep{cairns1987comparison}, so that for wolf pair $i$ and $j$ in year $t$,

$$
y_{ijt}\equiv\text{TWI}_{ijt} = \frac{N(\text{$i$, $j$ together in $t$})}{N(\text{$i$, $j$ together in $t$}) + N(\text{$i$ without $j$ in $t$}) + N(\text{$j$ without $i$ in $t$})},
$$

where $N(\cdot)$ denotes the number of occasions. 
\bigbreak
Our response data comprised 19,555 TWI observations. These data corresponded to 1061 wolf-years from 398 wolves between 1995 and 2024. Fifty seven unique packs were identified during 1022 wolf-years. The remaining thirty nine wolf-years were removed from the analysis because the focal wolf was not associated with any pack. An additional four wolves (0209F, 0353F, 0471F, 1154F) were removed from the analysis due to year(s) with missing pack information. The network experienced substantial turnover during the three-decade study period as wolves were born and died and packs formed and dissolved. On average, an individual wolf was in the data set for 2.64 years (SD = 1.95) and a pack persisted for 4.70 years (SD = 5.76). Another feature of our data was the highly skewed TWI distribution (Figure \ref{hist}). Of all pairwise observations, 88\% were exactly zero. The remaining 12\% consisted of continuous proportions (11\%) and exact ones (1\%). Such zero inflation presents well-known challenges for regression-based modeling, particularly when the zeros are structural instead of arising from sampling processes \citep{warton2005many}. Notably, 99\% of the zero TWIs occurred between wolves of different packs. The median inter-pack TWI was 0 (SD = 0.03) and the median intra-pack TWI was 0.71 (SD = 0.29). This pattern aligns with the well-documented social structure of gray wolves, which live in cohesive, territorial packs where interactions occur predominantly among pack members \citep{mech2019wolves}. The biological understanding of wolf social structure and the quantitative evidence from the data motivated the application of our MS-DLSM, with groups defined by pack membership. Because the TWI distribution consisted of exact zeros and ones and continuous proportions between zero and one, we modified the data model in Section \ref{data} to accommodate this mixture of two data types,
\begin{align*}
    \begin{cases}
        &y_{ijt}\left|p_{ijt}\right. \sim \text{Bern}(p_{ijt}),\ \text{if } y_{ijt}\in\{0,\ 1\}\\
        &\text{logit}(y_{ijt})\left|\mu_{ijt}, \sigma^2\right. \sim \text{N}\left(\mu_{ijt},\  \sigma^2\right),\ \text{if } y_{ijt}\in(0, 1)
    \end{cases},
\end{align*}
where
\begin{align*}
    \text{logit}(p_{ijt}) &= \mu_{ijt} = \bm{x}_{ijt}'\bm{\beta} - \log(d_{ijt}).
\end{align*}

In addition to pack membership, we incorporated three explanatory variables to capture ecological, demographic, and social drivers of interaction patterns: prey abundance in year $t$, an indicator for whether wolves $i$ and $j$ were the same sex, and the average pack size of the pair in year $t$. Elk (\textit{Cervus canadensis}) are the primary prey for gray wolves. Annual variation in elk abundance may influence overall network density, as cooperative behavior becomes increasingly important when prey are scarce \citep{metz2012seasonal, metz2020wolves}. Sex composition may also structure interactions. Wolf packs are typically organized around a dominant breeding pair composed of opposite-sex individuals, and this pair often maintains stronger cooperative bonds relative to other pairs \citep{peterson2002leadership}. Finally, pack size may influence interaction frequency. Members of smaller packs tend to associate more frequently on a per-dyad basis than members of larger packs, where social interactions are distributed across more individuals \citep{benson2025intrinsic}. We standardized elk abundance and pack size to have mean zero and variance one. Pairwise Pearson's correlations among all explanatory variables were below 0.2, precluding the concern on multicollinearity.
\bigbreak
We implemented the MS-DLSM for gray wolves in R \citep{R} using an MCMC algorithm. The chain was run for 200,000 iterations, with the first 80,000 iterations discarded as burn-in. The remaining samples were thinned at a rate of 1/10. All population parameters (Table \ref{post}) converged with potential scale reduction factors less than 1.1 \citep{gelman1992inference}. For comparison, we fitted two naive DLSMs: one with the same explanatory variables as the MS-DLSM, and a second that additionally included an indicator for whether wolves $i$ and $j$ were the same pack in year $t$. Because pack membership was already modeled via random effects in the MS-DLSM, including it again as a fixed effect could cause identifiability issues. Consequently, the first naive DLSM served as a benchmark for the regression structure, whereas the second served as a benchmark for the information incorporated.

\subsection{Results}
The MS-DLSM produced biologically plausible inferences on wolf interactions (Table \ref{post}). As expected, lower elk abundance was associated with increased interactions ($\beta_1$), opposite-sex pairs exhibited stronger interactions than same-sex pairs ($\beta_2$), and members of smaller packs interacted more frequently than those of larger packs ($\beta_3$). These effects were consistent in direction and statistical significance with the second naive DLSM. The second naive model further indicated that pairs within the same pack had the strongest effect on interaction strength, even after accounting for the other drivers, thereby providing retrospective support to the mechanistic specification of our MS-DLSM. In contrast, the first naive DLSM failed to detect a significant effect of prey abundance and inferred a counterintuitive effect of pack size. Moreover, the first naive model performed substantially worse than the other two models based on both CRPS and DIC. While the MS-DLSM and the second naive DLSM yielded qualitatively similar inferences, the MS-DLSM demonstrated a superior performance through a 23\% reduction in CRPS and a 22\% reduction in DIC.  
\bigbreak
We specified informative priors for the latent space movement models ($\sigma^2_{\mu_0}, \sigma^2_\mu, \sigma^2_{z_0}, \sigma^2_z$) to address the challenge posed by the high proportion of zero interactions, particularly between packs. As a result, the marginal posterior distributions of these parameters were concentrated near their prior means. Figure \ref{map} illustrates the latent space inference from the MS-DLSM. The left panel shows posterior mean latent trajectories of 10 unique packs from 1995 to 1999. Packs initiated near the origin of the latent space and gradually dispersed outward, creating separation in social space. Consequently, established packs occupied peripheral positions and newly formed packs occupied more central positions in a population social structure. The right panel focuses on the latent trajectories of 13 individuals over the same period from ``Mollie's" pack, originally named ``Crystal Creek" pack and the only pack remaining in Yellowstone whose direct origin is from the reintroduction. Unlike pack-level dynamics, individual wolves largely remained close to their latent movement center (rescaled to the origin following Equation \ref{i1t}), demonstrating strong adherence to their pack’s social core. This contrast between individual- and group-level social dynamics is consistent with empirical observations and would not have been illuminated under a naive DLSM or existing clustering DLSMs that ignore cluster scale movement. Post-hoc examination identified two ``social outliers": wolf 0003M, who in his only year of observation spent 72\% of interactions with the dominant pack and was associated with two different packs, and wolf 0118M, who in his second year spent 63\% of interactions with the dominant pack and was associated with three different packs. For comparison, an average wolf in our data spent almost all interactions with its dominant pack (mean = 0.95, SD = 0.12) and was rarely associated with more than one pack (mean = 1.26, SD = 0.59) in a year. Importantly, this information was not used in model fitting; nonetheless, the MS-DLSM was able to identify these social outliers through mechanistic inference.

\section{Discussion}\label{diss}
We extended existing DLSMs by introducing coupled movement models that separate group- and individual-level dynamics in clustering networks. We also developed an approximate decomposition of latent pairwise distances enabled by our multi-scale movement framework. Compared to a naive DLSM, the MS-DLSM improved inference and overall performance when clustering is strong. In addition, our model addressed challenges inherent to many social networks, like in gray wolves, characterized by rapid turnover, zero inflation, and a mixture of response data types. Although demonstrated for a social predator, our framework is widely applicable to various types of clustered networks, including other group-living wildlife species, workplace collaboration, urban gang networks, and military units.
\bigbreak
Beyond qualitative knowledge of a network’s propensity to cluster, a quantitative criterion for assessing whether clustering is sufficiently strong to favor the MS-DLSM over a naive-DLSM (i.e., whether $\frac{\sigma^2_\mu}{\sigma^2_z}\gg 1$) is the proportion of variance explained by a “same group” indicator. For example, we defined an explanatory variable $\bm{W}$ with $w_{ijt} = \mathbb{I}(c(i, t) = c(j, t))$ and regressed $\text{logit}(y_{ijt})$ on both $\bm{x}_{ijt}$ and $w_{ijt}$ using R's \texttt{lm()} function. The goal is \textit{not} to estimate the true $\beta$ coefficients, but to quantify the unique contribution of $\bm{W}$ in explaining the variation in response. We computed the semi-partial correlation squared as

$$
    sr^2_w = \frac{\text{RSS}_{\text{reduced}} - \text{RSS}_{\text{full}}}{\text{RSS}_\text{null}},
$$

where $\text{RSS}_{\text{full}}$, $\text{RSS}_\text{reduced}$, and $\text{RSS}_\text{null}$ were the residual sum of squares for the full model (with $\bm{X}$ and $\bm{W}$), the reduced model (with only $\bm{X}$), and an intercept-only model, respectively. For generalized linear models, we used \texttt{glm()} and deviance in place of residual sum of squares. In simulation, Case 1 ($\frac{\sigma^2_\mu}{\sigma^2_z} = 100$) yielded a $sr^2_w$ of 0.63, Case 2 ($\frac{\sigma^2_\mu}{\sigma^2_z} = 10$) of 0.36, and Case 3 ($\frac{\sigma^2_\mu}{\sigma^2_z} = 1$) of 0.15. In the case study, $sr^2_w$ (0.82) was higher than in Case 1, indicating strong clustering, consistent with improved performance of the MS-DLSM, and supporting our choice of informative priors.
\bigbreak
In simulation, we illustrated computational gains from the distance decomposition (Equation \ref{d}); however, the precise trade-off depends on the complexity of the data structure and the efficiency of implementation. In the case study, our MS-DLSM was slower than a naive DLSM due to the substantial computational cost of mapping dynamic pack memberships to irregularly sized groups. Performance could be improved through more efficient bookkeeping and implementation strategies in a faster programming environment. In particular, compiled approaches within R (e.g., \texttt{Rcpp}, \text{RStan}) or alternative languages with C++ integration (e.g., Julia) may substantially reduce runtime. Although the algorithm could be parallelized across groups, we did not pursue this because thread-management overhead (e.g., via \texttt{doParallel}) outweighed the benefits; nonetheless, parallelization remains a promising approach as group-wise computations grow more intensive.
\bigbreak
Although developed for undirected networks, our framework can be readily extended to directed networks, allowing differentiation of in-degree versus out-degree and the study of individual popularity (high in-degree) versus sociality (high out-degree). These individual-specific parameters can be incorporated as additive random effects (e.g., \cite{krivistky2009representing}):
\begin{align*}
        \text{logit}(y_{ijt}) &= \bm{x}_{ijt}'\bm{\beta} - \log(d_{ijt}) + \delta_{i, t} + \gamma_{j, t} + \epsilon_{ijt},\\
        \delta_{i, t} &\sim \text{N}\left(0, \sigma^2_\delta\right),\ \gamma_{j, t} \sim \text{N}\left(0, \sigma^2_\gamma\right);
\end{align*}
or as multiplicative random effects on latent distance (e.g., \cite{sewell2015latent}):
\begin{align*}
        \text{logit}(y_{ijt}) &= \bm{x}_{ijt}'\bm{\beta} - \frac{\log(d_{ijt})}{r_i} - \frac{\log(d_{ijt})}{r_j} + \epsilon_{ijt},\\
        (r_1, \dots, r_n)' &\sim \text{Dirichlet}(\bm{\alpha}).
\end{align*} 
A natural extension is to allow mixed group memberships \citep{xing2010a}, which is relevant for wolves that associate with multiple packs in a given year. A latent movement center accommodating cluster mixing can be expressed as

$$
    \bm{\mu}(c(i, t)) = \sum_{k = 1}^K p_k(i, t) \bm{\mu}(k,t),\ p_k(i, t) > 0,\ \sum_{k = 1}^Kp_k(i, t) = 1,
$$

where $\bm{\mu}(k, t)$ represents the social core of pack $k$ in year $t$ and $p_k(i, t)$ the proportion of time wolf $i$ spends with pack $k$. Distance decomposition (Equation \ref{d}) may not apply under a mixture framework; however, the coupled movement models would still provide valuable inference. Further, our framework can be integrated with existing DLSMs that cluster individuals via hidden Markov models (e.g., \cite{daniel2023bayesian}). This would allow us to impute missing group memberships and forecast interactions of individuals whose memberships change over time.

\section*{Acknowledgment}
We thank Daniel Stahler, Erin Stahler, Kira Cassidy, Douglas Smith, pilots Mark Packila and Roger Stradley, and numerous technicians for their assistance with data collection and curation. We also thank Andee Kaplan and Mevin Hooten for their feedback on the writing and organization of the manuscript.

\newpage
\begin{nolinenumbers}
\bibliographystyle{biom.bst} 
\bibliography{refs}

\section*{Supplementary Materials}
Web Appendices are available with this paper at the Biometrics website on Oxford Academic.

\newpage
\section*{Figures}
\begin{figure}[H]
    \centering
    \includegraphics[scale=1, trim=5mm 20mm 0mm 20mm]{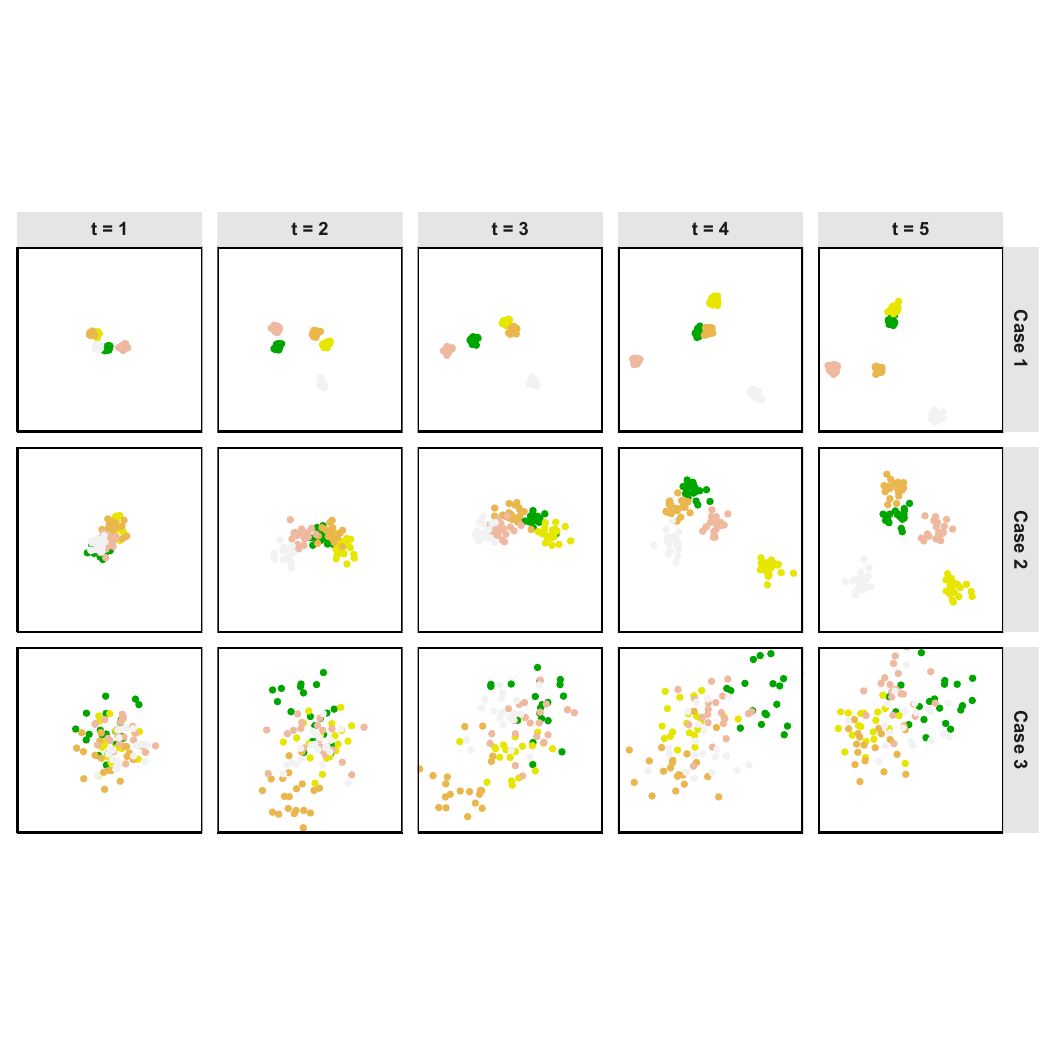}
    \caption{Simulated trajectories from one hundred nodes and five color-coded groups in a two-dimensional latent space. Rows represent cases. Columns represent times. In all three cases, the nodes remain close to their respective movement centers according to the discrete time Ornstein-Uhlenbeck model, and the clusters become dispersed over time according to the random walk model.}
    \label{sim_ts}
\end{figure}

\newpage
\begin{figure}[H]
    \centering
    \includegraphics[scale=0.75, trim=0mm 0mm 0mm 5mm]{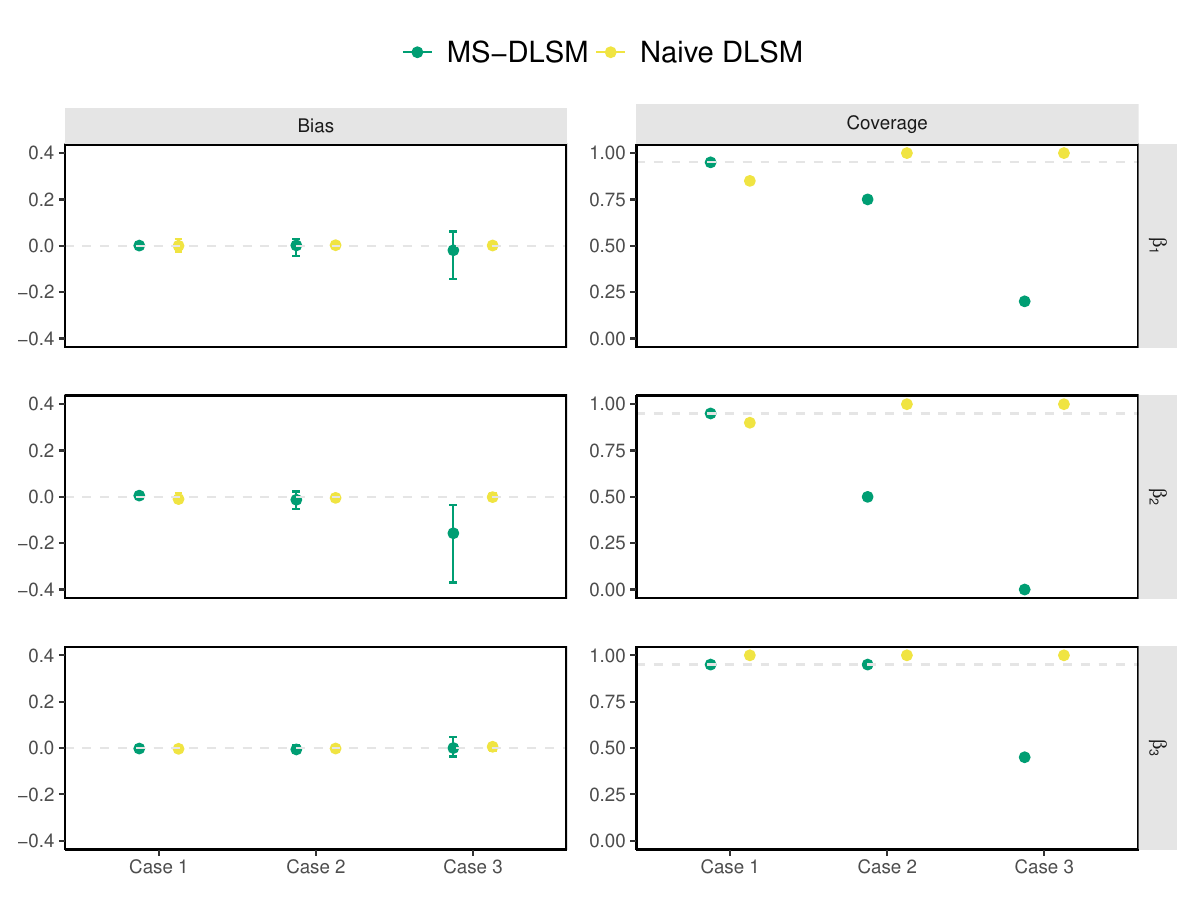}
    \caption{A comparison of inference quality between the MS-DLSM and the naive DLSM. For bias, points represent the average metric and error bars represent the 95\% interval from the twenty simulated data sets. For coverage, a point metric was summarized from all simulations under each case.}
    \label{sim_infer}
\end{figure}

\newpage
\begin{figure}[H]
    \centering
    \includegraphics[scale=0.75, trim=0mm 0mm 0mm 20mm]{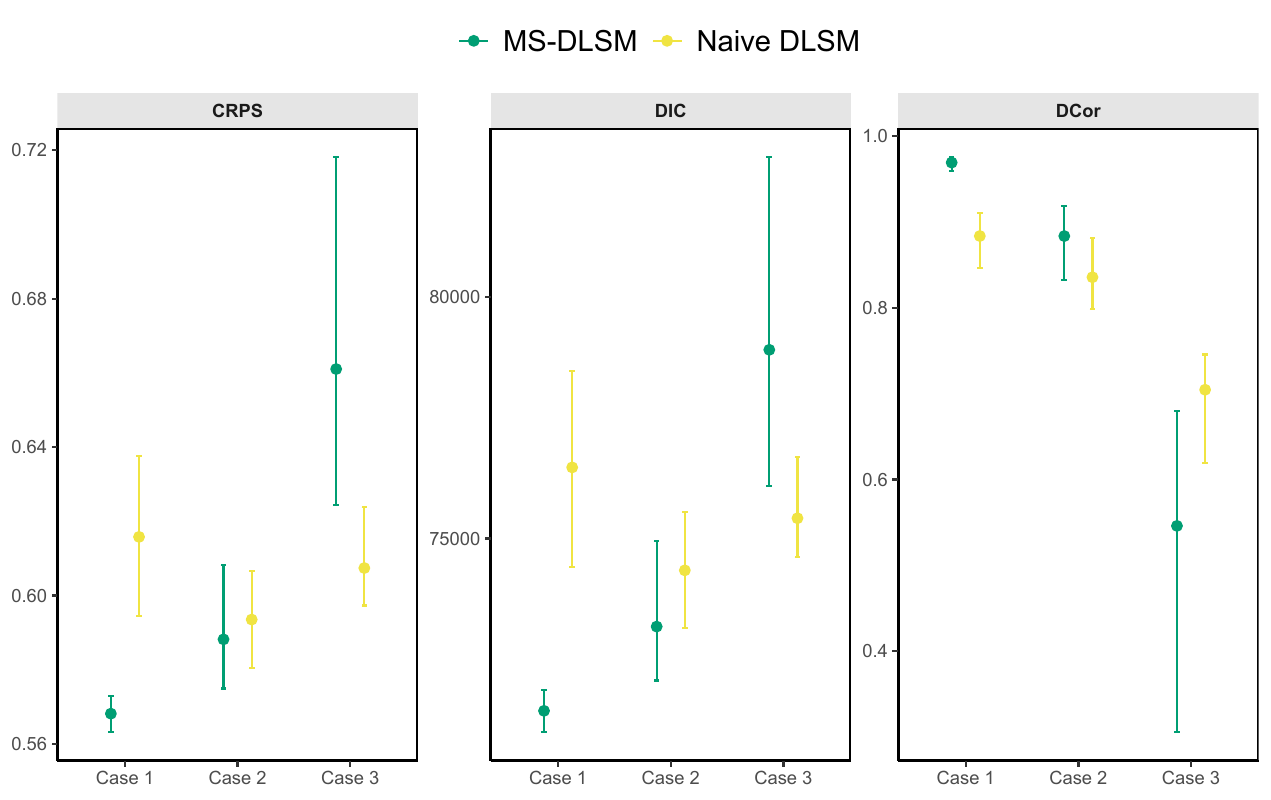}
    \caption{A comparison of overall model performance between the MS-DLSM and the naive DLSM. Points represent the average metric and error bars represent the 95\% interval from the twenty simulated data sets under each case.}
    \label{sim_fit}
\end{figure}

\newpage
\begin{figure}[H]
    \centering
    \includegraphics[scale=0.75, trim=0mm 5mm 0mm 10mm]{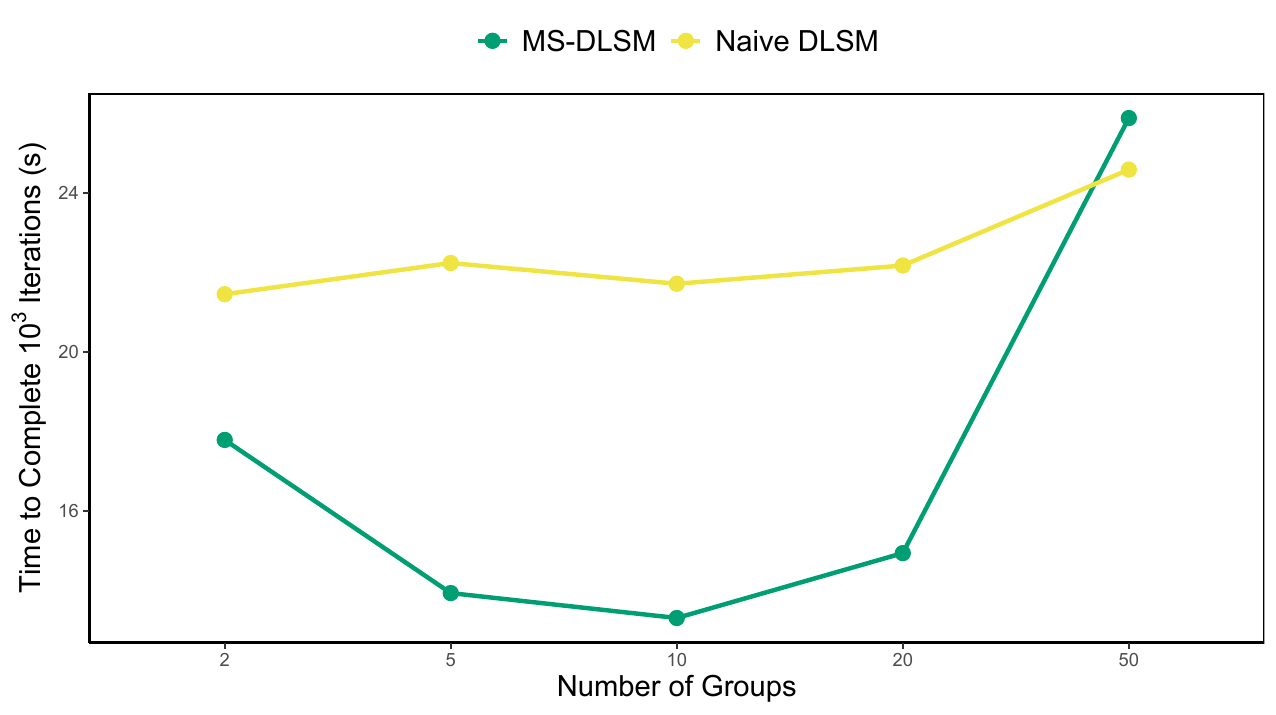}
    \caption{A comparison of computational cost between the MS-DLSM and the naive DLSM under various numbers of groups.}
    \label{time}
\end{figure}

\newpage
\begin{figure}[H]
    \centering
    \includegraphics[scale=0.75]{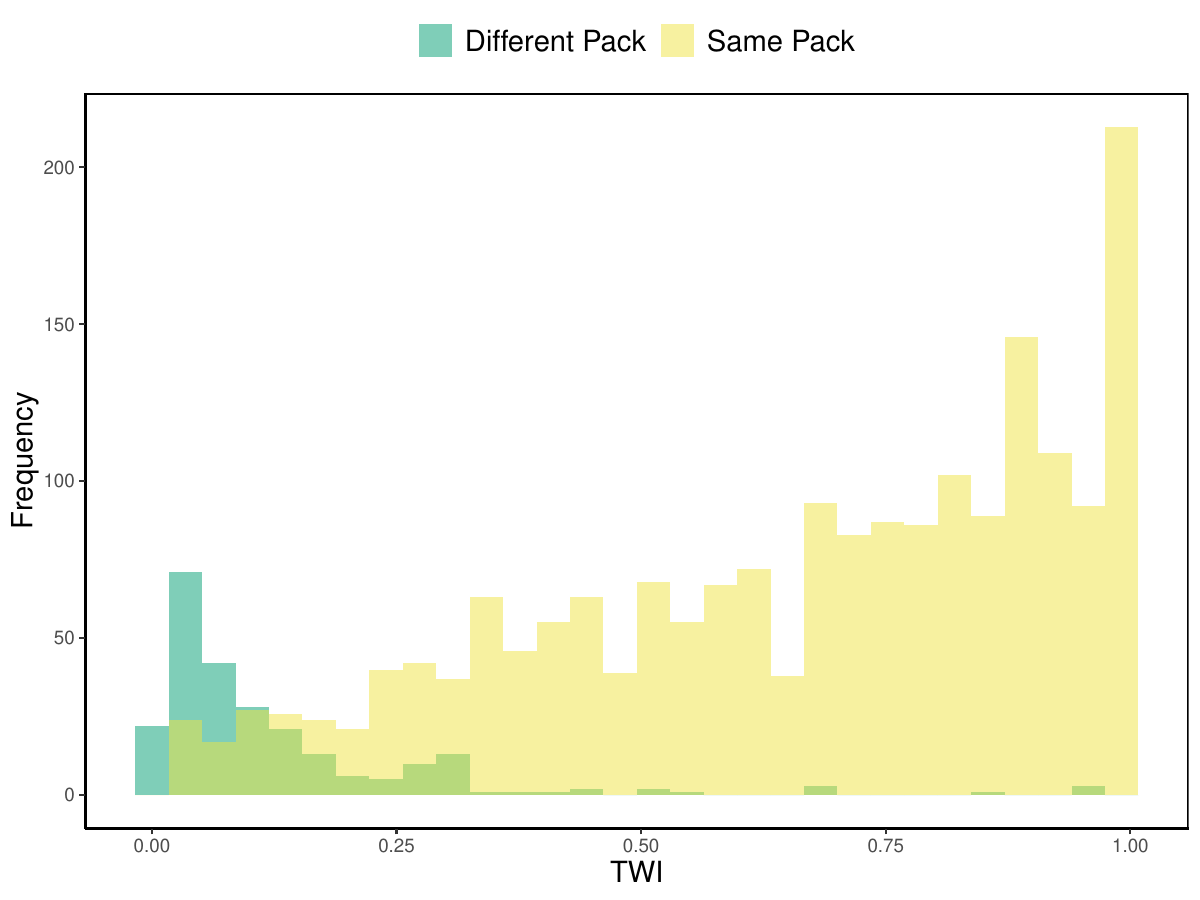}
    \caption{TWI distribution after removing zeros. On average, the inter-pack TWI was much lower than the intra-pack TWI.}
    \label{hist}
\end{figure}

\newpage
\begin{figure}[H]
    \centering
    \includegraphics[scale=0.85, trim=10mm 10mm 0mm 20mm]{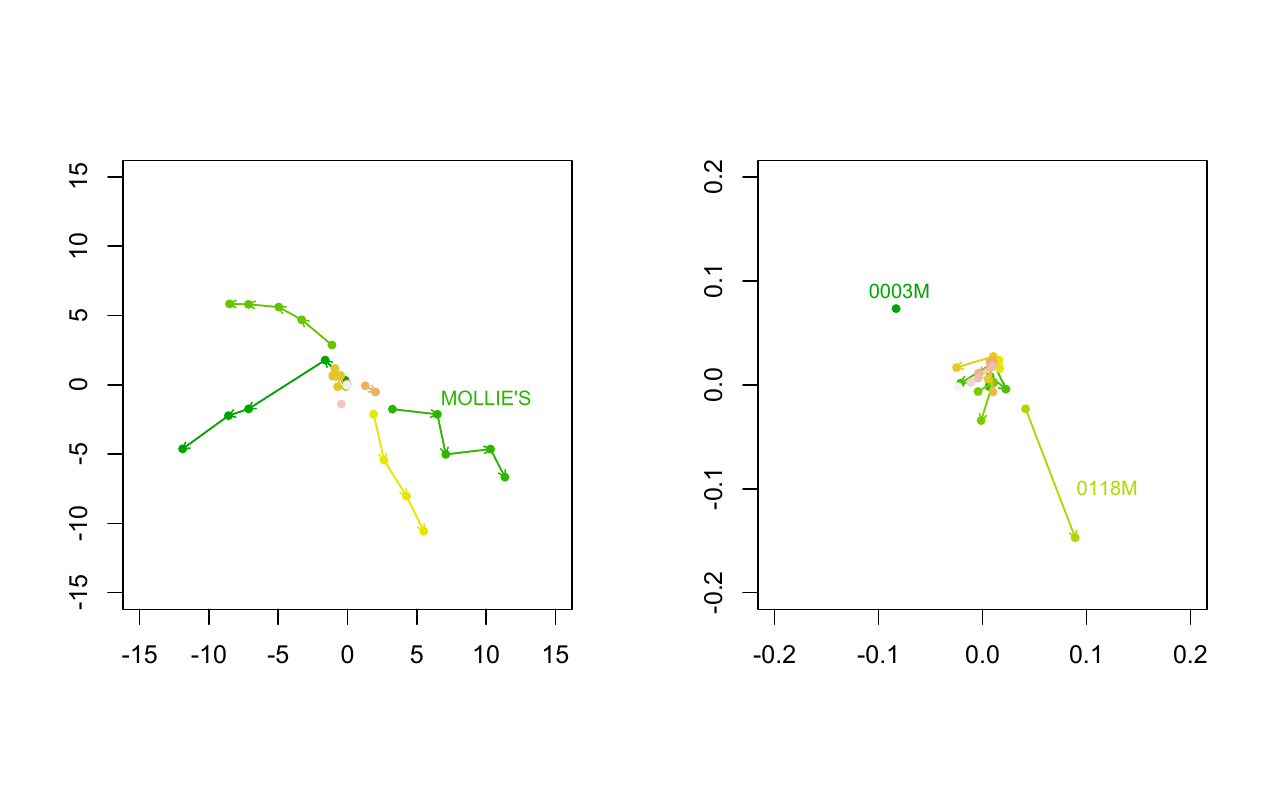}
    \caption{An example of posterior mean latent trajectories at the group (left) and individual (right) levels in the case study. The colors represent different packs (left) or wolves (right). The color gradient represents the sequence of entry into the network, with green representing earlier entries and pink representing more recent ones. The arrows show the direction of movement. While the packs (left) dispersed in the social space over time, most wolves (right) adhered to their pack's social core.}
    \label{map}
\end{figure}

\newpage
\section*{Tables}
\begin{table}[H]
    \centering
    \begin{tabular}{lccc}
        Parameter/Metric & MS-DLSM & Naive DLSM & Naive DLSM + Same Pack \\\hline
        $\beta_1$ (Prey Abundance) & -0.17 (-0.25, -0.10) & 0.01 (-0.06, 0.08) & -0.11 (-0.17, -0.04) \\
        $\beta_2$ (Same Sex) & -0.92 (-1.00, -0.84) & -1.58 (-1.70, -1.45) & -2.28 (-2.39, -2.17) \\
        $\beta_3$ (Pack Size) & -0.10 (-0.17, -0.04) & 0.14 (0.08, 0.21) & -0.15 (-0.21, -0.10)\\
        $\beta$ (Same Pack) & - & - & 3.25 (3.15, 3.35)\\
        $\sigma^2$ (Var. TWI) & 1.51 (1.39, 1.64) & 6.38 (5.67, 7.07) & 2.13 (1.96, 2.30) \\
        $\sigma^2_{\mu_0}$ (Init. Var. Pack) & 1.03 (0.99, 1.07) & - & - \\
        $\sigma^2_\mu$ (Var. Pack) & 1.08 (1.04, 1.13) & - & - \\
        $\sigma^2_{z_0}$ (Init. Var. Ind.) & 0.02 (0.02, 0.02) & 1.19 (1.14, 1.24) & 1.23 (1.19, 1.28) \\
        $\sigma^2_z$ (Var. Ind.) & 0.02 (0.02, 0.02) & 1.04 (1.00, 1.08) & 1.07 (1.03, 1.12) \\
        $\rho$ (Autocorr. Ind.) & 0.95 (0.88, 1.00) & 1.00 (1.00, 1.00) & 1.00 (1.00, 1.00) \\
        \hline
        CRPS & 0.08 & 0.19 & 0.11 \\
        DIC & 8442 & 14247 & 10764 \\
        \hline
    \end{tabular}
    \caption{A comparison of posterior inference (posterior mean and 95\% credible interval of respective marginal distributions) and overall performance among the three models in the case study.}
    \label{post}
\end{table}

\end{nolinenumbers}

\end{document}